\documentclass[aps,preprint]{revtex4}
\usepackage{amsmath}
\usepackage{graphicx}
\begin{document}
\title{Quasiparticle Resonances in the BCS Approach} 
\author{R. Id Betan}
\affiliation{\small\it Departamento de Fisica y Q\'imica, FCEIA, UNR,
 Avenida Pellegrini 250, 2000 Rosario, Argentina}
\email{idbetan@ifir.edu.ar},
\author{N. Sandulescu}
\affiliation{\small\it Institute of Physics and Nuclear Engineering,
76900 Bucharest, Romania\\
Institut de Physique Nucleaire, Universite Paris-Sud, 91406 Orsay Cedex, France}
\email{sandules@ipno.in2p3.fr},
\author{T. Vertse}
\affiliation{\small\it Institute of Nuclear Research of the
Hungarian  Academy of Sciences,
H-4001 Debrecen, Pf. 51, Hungary\\
University of Debrecen, Faculty of Information Science,H-4010 Debrecen, Pf. 12, Hungary }
\email{vertse@tigris.klte.hu},
\date{\today}
\begin{abstract}
We present a simple method for calculating the energies and the widths of
quasiparticle resonant states. The method is based on BCS equations solved
in the Berggren representation. In this representation the quasiparticle 
resonances are associated to the Gamow states of the mean field. The method is
illustrated for the case of neutron-rich nuclei $^{20-22}$O and $^{84}$Ni.
 It is shown that the contribution of the continuum
coupling to the pairing correlations is small and largely dominated by
 a few resonant states close to the continuum threshold.

\end{abstract}

\maketitle

\section{Introduction}

In nuclei close to the drip lines the low-lying excited states belong mostly
to the continuum part of the spectrum. It is therefore not surprising that
a lot of effort has been devoted recently to the treatment of the continuum in
nuclear models.

One of the first attemps to evaluate continuum configurations
in nuclei was done within the  continuum shell-model ( see 
Ref.\cite{zel} and the references therein). In this shell model approach
the coupling to the single-particle continuum is introduced through the 
scattering states of real energies. A different approach for evaluating the
continuum processes in the shell-model framework is based on Berggren
representation. In this representation the single-particle continuum
is formed by a finite number of Gamow resonances and a continous set
of scattering states of complex energy. The advantage of a shell model
based on Berggren representation is that it allows a direct calculation
of multi-particle resonant states in nuclear systems \cite{cxsm1,cxsm2,
cxsm3,pol,naz}.

The continuum part of the excitation spectrum was also investigated
using mean field techniques. One of the most known
of these approaches is the continuum-RPA \cite{bertsch,giai}.
In this approach, well suited for the description of particle-hole ($p-h$)
excitations
in closed shell nuclei, the contribution of the continuum to the $p-h$ response
function is calculated exactly. To single out from the continuum $p-h$ response the 
states corresponding to the $p-h$ resonances, a resonant-RPA approach was also
developed
\cite{ver,cur}, in which the coupling to the single-particle continuum is introduced
through a finite set of Gamow functions.

For even-even nuclei with open shells, where pairing correlations are important,
the continuum part of the spectrum corresponds to two-quasiparticle excitations.
The unbound two-quasiparticle excitations  can be described in the continuum-QRPA
approach \cite{matsuo,khan}. Self-consistent continuum-QRPA calculations are
presently done only with zero-range forces \cite{khan}. At variance with the
continuum-RPA, in these calculations the continuum contribution is taken
into account only up to an energy cut-off. 

 In odd-even nuclei the most simple excitations are of the one quasiparticle
 type. The continuum part of the one quasiparticle spectrum can be calculated 
 by solving the HFB equations with scattering type boundary conditions
 \cite{grasso}. By using the behaviour of the phase shift  close to a resonance,
 one can find the quasiparticle resonances of physical interest. The
 quasiparticle resonances can be found also by solving the HFB equations in
 complex energy plane, as have been already done for some particular
 hamiltonians \cite{belyaev}.

 However, the simplest framework to evaluate one quasiparticle resonances
 is the BCS approach. The contribution of resonant states to pairing
 correlations can be introduced in the BCS equations
 through the scattering states with energies located in the vicinity of
 single-particle resonance energies \cite{ni1,ni2,ni3}. Alternatively, the
resonant continuum
 was studied by solving the BCS equations in a basis containing only 
 bound and Gamow states \cite{kruppa,clv}. The drawback of this study
 is that the global quantities associated to the bound state
 of a nucleus ( e.g., binding energy, radii)  become complex. The scope of
 this paper is to show how one can calculate within the BCS approach the
 energies and the
 widths of quasiparticle resonances  by taking into account the whole
 contribution of  single-particle continuum  to pairing correlations and,
 in the same time,  by preserving the correct properties of physical
 observables.

\section{Formalism}
\label{sec:form}
\subsection{Berggren representation}

 The quasiparticle spectrum is described here within the  Berggren representation.
 Since this representation is not very common in 
 nuclear structure calculations, we will first describe it briefly, emphasizing 
 those features that we will need in the present paper.
 For more details see Refs. \cite{tor,bl,lio}. 

 The Berggren representation is formed by bound states, Gamow resonances, and a 
 continuous set of scattering states of complex energy. For a given mean
 field potential (e.g., of Woods-Saxon type), the Gamow states are the
 outgoing solutions of the Schrodinger equation and correspond to a discrete
 set of complex wave numbers $k_\nu=\kappa_\nu -i\gamma_\nu$. To each Gamow
 state $u_{\nu lj}(r)$ with the wave number $k_\nu=\kappa_\nu -i\gamma_\nu$ is
 associated a "mirror"  state $\tilde{u}_{\nu lj}(r)$ with the wave number
 $\tilde{k}_\nu=-\kappa_\nu -i\gamma_\nu$, 
 such that $\tilde{u}^*_{\nu lj}(r) \equiv u_{\nu lj}(r) $. The Gamow states
and
 their 
 mirrors form a biorthogonal set which can be normalized to unity by using 
 various regularization schemes \cite{tor,vertse}. All the wave functions
 written above are the radial single-particle wave functions of angular
 momentum ${lj}$.

\begin{figure}
       \includegraphics[width=0.5\textwidth]{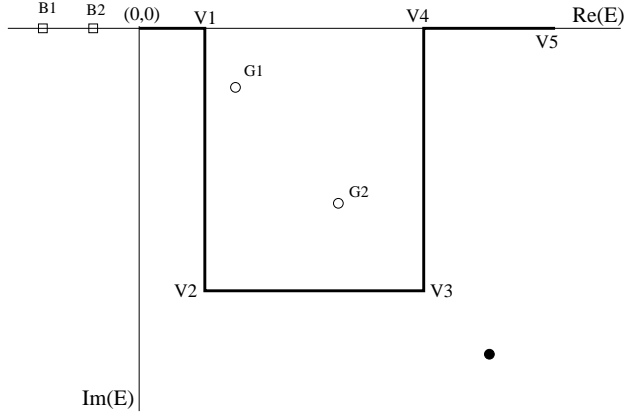}
 \caption{Contour on the complex energy plane representing the continuum path.
          The points $V_i$ are the vertices defining the contour.
          The open circles labelled by $G_i$ indicate the complex energy of the Gamow resonances enclosed by the contour while the open squares label by $B_i$ are the bound states. The dot is a pole of the S-matrix excluded from the Berggren expansion\label{fig:cont}}.
 \end{figure}

 The scattering states introduced in the Berggren representation belongs to 
 a contour in the complex momentum or complex energy plane. An example of such a
 contour
 in the complex energy plane is shown in Figure \ref{fig:cont}.

The contour encloses the poles 
 associated to the  Gamow states which are taken explicitely into the Berggren
 representation. The scattering states belonging to the contour, 
 $u_{lj}(\varepsilon,r)$, the enclosed Gamow resonances, $u_{\nu lj}(r)$, 
 and the bound states, $u_{nlj}(r)$, form a complete set, i.e.,
 \begin{equation}\label{eq:del}
    \delta(r-r')=\sum_n\; u_{nlj}(r)\; u_{nlj}(r') + 
                 \sum_\nu\; u_{\nu lj}(r)\;u_{\nu lj}(r')+
                 \int_L\; d\varepsilon\; u_{lj}(\varepsilon,r)\;
                                         u_{lj}(\varepsilon,r').
 \end{equation}
 In numerical applications the integral along the path $L$ is evaluated by
 using a finite number of scattering states of energy $\varepsilon_i$
 defined by a chosen quadrature rule. In what follows this set of scattering
 wave functions together with the bound and Gamow  states are denoted by a 
 unique function, i.e.,

 \begin{displaymath}\label{eq:dis}
    \psi_{ilj}(r)= \left\{
                   \begin{array}{ll}
                    u_{ilj}(r) & \textrm{bound states and  resonances} \\
      \sqrt{\chi_i} u_{lj}(\varepsilon_i,r) & \textrm{scattering states}~,
                        \end{array}
                 \right.
 \end{displaymath}
where the quantities $\chi_i$  depends on the quadrature. In general, they
are defined by $\chi_i=\omega_i \dot{L}_i$, where $\omega_i$ are the quadrature 
weights and $\dot{L}_i$ are the derivative of the complex contour with respect
to the parametrisation variable. 

In principle, a Berggren representation can be formed by using any set of
Gamow
states enclosed by a contour in the complex plane. However, from physical
point of view it is better to choose a basis which includes explicitely
only narrow Gamow resonances. The contribution of wide resonances which are
not included explicitetly in the Berggren representation 
is automatically taken into account through the scattering states belonging
to the contour.

\subsection{BCS equations in Berggren representation}

 The BCS equations in Berggren basis will be derived here from the
 Hartree-Fock-Bogolibov (HFB) equations in coordinate representation. 
 For zero range forces and for spherically symmetric systems the radial HFB 
 equations have the following form  \cite{deGennes}:
 \begin{equation}
  \begin{array}{c}
    \left( \begin{array}{cc} h(r) - \lambda & \Delta(r) \\
           \Delta(r) & -h(r) + \lambda \end{array}
    \right)
    \left( \begin{array}{c} U_k(r) \\  V_k (r) \end{array}
    \right) = E_k
    \left( \begin{array}{c} U_k (r) \\  V_k (r) \end{array} \right),
  \end{array}
\end{equation}
where $\lambda$ is the chemical potential, $h(r)$ and $\Delta(r)$ are the
mean field hamiltonian and pairing field, respectively.
$U_k(r)$ and $V_k(r)$ are the up and down components of the HFB wave function,
denoted below by $|\Psi_k>$).
 At large distances $U_k(r)$ decays exponentially for $E_k < -\lambda$ and
 behaves as a scattering state for $ E_k> -\lambda$.
 On the other hand, the function $V_k(r)$ goes to zero exponentially for any
 value of the quasiparticle  energy (for a general discussion on the asymtotic
 properties of the HFB wave  functions see \cite{bulgac}).
  Consequently, the quasiparticle spectrum is discrete for $E_k < -\lambda$ 
and continous  for $E_k > -\lambda$. In the ground state of the system the
particle density and the pairing density are given by:

\begin{equation} \label{2}
      \rho(r) =\frac{1}{4\; \pi} \sum_k\; (2j_k+1)\; V_k^*(r)\; V_k(r)~~~.
 \end{equation}

 \begin{equation} \label{3}
    \tilde{\rho}(r) = \frac{1}{4\; \pi} \sum_k\; (2j_k+1)\; U_k^*(r)\;
V_k(r)~~~.
  \end{equation}
 The summations in this equation is over the quasiparticle
 spectrum of the system (for $E_k > -\lambda$ the sum should be replaced 
 by an integral) up to a given cut-off energy.

 In the calculations presented in the next section we take for the mean field
 a Wood-Saxon potential. The pairing field is calculated self-consistently
 by using a pairing interaction of the following form \cite{be}:

\begin{equation}
V (\mathbf{r}-\mathbf{r^\prime}) = V_0 [1 -
\eta(\frac{\rho}{\rho_0})^{\alpha}]
\delta(\mathbf{r}-\mathbf{r^\prime})
\equiv V_{eff} \delta(\mathbf{r}-\mathbf{r^\prime}).
\end{equation}
With this pairing interaction the pairing field is given by:
\begin{equation}
\Delta(r) = \frac{V_{eff}}{2} \tilde{\rho}(r).
\end{equation}

 Before going further with the derivation, it is worth mentioning how the
particle and pairing densities behave 
 in excited states belonging to the continuum part of the quasiparticle
 sectrum of odd-even nuclei. Formally, in order to get the excited state coresponding to a state
 $k$ of an odd-even nucleus one can simply interchange the eigenvector $(U_k, V_k)$ by
 $(V_k^*,U_k^*)$ \cite{rs}.  Consequently, the particle density corresponding to an
 excited state $k$ will depend on $|U_k|^2$. Thus, if the excitated state belongs
to the continuum, the function $U_k$ is of a scattering type; hence,  there
 is a finite probability for the particle to be found at infinity, as physically required.
 On the other hand, by interchanging $U_k$ and $V_k$ one does not affect the 
 asymtotic behaviour of pairing density,  which is still going to zero exponentially  at very large
 distances. This is again a physically meaningful condition since when a
particle is very far from the nucleus, its contribution to the pairing
 correlations should vanish.

 Starting from the HFB equations in coordinate space we can easily get the HFB equations
 in  Berggren representation. In order to accomplish that, we first expand the HFB wave
 functions in the Berggren basis:

 \begin{equation} \label{eq:expu}
      U_k(r) = \sum_i\; <\tilde{\psi}_i|U_k>\; \psi_i(r)
        \equiv \sum_i\; U_{i,k}\; \psi_i(r)
 \end{equation}
 \begin{equation} \label{eq:expv}
      V_k(r) = \sum_i\; <\tilde{\psi}_i|V_k>\; \psi_i(r)
        \equiv \sum_i\; V_{i,k}\; \psi_i(r)~~,
 \end{equation}
where the expansion coeficients, related to the Bogoliubov transformation, are complex quantities.
 Then one multiplies the HFB equations from the left with the complex conjugate of a mirror Berggren
 vector, $\tilde{\psi}_i^*(r)$ and one performs the radial integration.
 One thus gets the HFB equations in  Berggren representation:

 \begin{equation} \label{eq:hfbm}
     \begin{array}{c}
        \left(
        \begin{array}{cc}
           \varepsilon_i - \lambda & \Delta_{i,j} \\
           \Delta_{i,j} & -\varepsilon_i + \lambda
        \end{array}
        \right)
        \left(
        \begin{array}{c}
            U_{i,j} \\
            V_{i,j}
        \end{array}
        \right ) = E_i
        \left(
        \begin{array}{c}
            U_{i,j} \\
            V_{i,j}
        \end{array}
        \right),
     \end{array}
 \end{equation}
which formally have the same structure as in any hermitic representation.
 The essential difference is that the matrix elements of the pairing field are defined with the Berggren metric, i.e., by using mirror states in the bra positions and employing regularisation techniques for calculating the diverging integrals.
 Thus, $\Delta_{i,j}$ is defined by:
 \begin{equation} \label{eq:pad1}
      \Delta_{i,j} = <\tilde{\psi}_i | \Delta(r) |\psi_j >
                   = \int dr r^2 \psi_i(r) \Delta(r) \psi_j(r)~,
 \end{equation}
where $\tilde{\psi}_i$ is the mirror state corresponding to $\psi_i$.

 Finally, from Eq. (\ref{eq:hfbm}) one can get the BCS equations in the Berggren
representation
 by neglecting the off-diagonal matrix elements of the pairing gap matrix.
 Physically, this approximation means that one neglects the pairing correlations associated to the 
 Cooper pairs formed in states which are not time-reversed.
 In this approximation the expansions (7,8) of the HFB wave
 function are reduced to one term.
 The corresponding expansion coeficients, denoted below by $u_i$ and $v_i$, 
 are complex quantities.
 They satisfy the condition $u_i^2 + v_i^2=1$, which is obtained from the Berggren
 normalization 
 condition of the HFB wave function  $\Psi_i$, i.e., $<\tilde{\Psi}_i |\Psi_i >=1$.
 By combining this condition with Eq.(\ref{eq:hfbm}) one gets the BCS 
 equations in the Berggren representation:

 \begin{eqnarray}
       v_i^2 & = &  \frac{1}{2}(1- \frac{\varepsilon_i-\lambda}{E_i}) \\
       E_i & = & \sqrt{(\varepsilon_i-\lambda)^2 + \Delta_i^2} 
\end{eqnarray}
where 
\begin{equation}
\Delta_i = \int\; dr\; r^2 \psi^2_i(r) \Delta(r)
\end{equation}
and the chemical potential is found from the particle number 
equation:
\begin{equation}
  N  =  \int \left[ \sum_i\; (2j_i+1)\; v_i^2\; \psi_i^2(r)
                        \right ] r^2 dr~~. 
\end{equation}

 In conclusion, apart from the radial integrals which are calculated with
 the Berggren metric (and a given regularisation procedure),  the BCS equations
 in the Berggren representation have  formally the same expression as in any
 hermitic and discrete basis. However, one should keep in mind  that the
 pairing gaps and the occupation probabilities associated to the scattering
 states correspond to an energy interval.

 The advantage of working in Berggren representation is that the quasiparticle
resonances appear as unique
 states, associated to the Gamow functions. In this way the widths of
 the quasiparticle resonances, given by the imaginary parts of the
 quasiparticle energies, are calculated unambiguously.

 In principle, one can get the complex energies associated with the resonant
 states by solving the BCS equations in complex energy plane.
 However, due to the strong non-linearity of the BCS equations, this is a difficult numerical task.
 One can avoid this task by using the fact that the pairing field and the chemical 
 potential do not depend on the representation.
 Thus, they can be calculated by solving the BCS equations in a real
 energy representation,  which can be obtained by choosing the $L$ contour
 to be the real energy axis.
 From the pairing field calculated using the real energy axis, we can then
 get, by using Eq.(11c), the pairing gaps associated to the 
 Gamow states. With these values of the pairing gaps, which are complex
 quantities, one can finally calculate, by using Eq.(11b),
 the energies of the resonant quasiparticle states.

\section{Applications} \label{sec:appl}

 The scope of the calculations presented below is twofold: a) to illustrate,
 within the approximation discussed above, how the energies and the widths
 of quasiparticle resonant  states behaves in neutron-rich nuclei;
 b) to analyse what is the relative contribution of resonant
    and non-resonant continuum  upon pairing correlations
    in nuclei close to the dripline.
 For these purposes we have performed  calculations for the isotopes
 $^{20-22}$O and $^{84}$Ni.

 \subsection{ Mean fields and single-particle states}

 The BCS calculations presented in this section are based on a mean
 field of Woods-Saxon form. In addition to  the mean field, we take
 also a  standard spin-orbit interaction, with the form factor given by the
 derivative of a Wood-Saxon function.
 The Wood-Saxon parameters we used for oxygen isotopes have the
 following values: $V_0$=55.8 MeV, 
 $V_{so}$=12.12 MeV, $r_0$=1.21 fm and  $a$ = 0.65 fm. The values of 
 $r_0$ and $a$ are the same for the mean field and for the spin-orbit
 interaction. With these parameters we get for the  isotope $^{17}$O 
 the single-particle energies shown in Table \ref{sps}.

 \begin{table}
 \caption{\label{sps}
           Single- particle energies (in MeV) corresponding to $^{17}$O and $^{79}$Ni}
 \begin{ruledtabular}
 \begin{tabular}{cccccc}
            & $^{17}$O &             &            & $^{79}$Ni & \\
  State     &         & Energies    & State      &           & Energies   \\
 \hline
  $1f_{7/2}$ &         & 7.45-i 1.53 & 1$h_{11/2}$ &          & 3.56-i 0.02 \\
  $1d_{3/2}$ &         & 0.91-i 0.05 & 1$g_{7/2}$  &          & 1.93-i 0.01 \\
  $2s_{1/2}$ &         &-3.26        & 2$d_{3/2}$  &          & 0.51-i 0.06 \\
  $1d_{5/2}$ &         &-4.12        & 3$s_{1/2}$  &          &-0.82        \\
  $1p_{1/2}$ &         &-14.57       & 2$d_{5/2}$  &          &-1.44        \\
 \end{tabular}
 \end{ruledtabular}
 \end{table}

 One can notice that
 the energy of the single-particle state $d_{3/2}$ is very close to the
 experimental value, equal to $(0.94 -i0.05)$ MeV \cite{exp} .
 The relevant single-particle resonances
 in oxygen isotopes are the states  $2d_{3/2}$ and $1f_{7/2}$.
 The corresponding complex energies shown in Table \ref{sps} are the energies
 of the Gamow states,i.e., $\varepsilon_\nu=\epsilon_\nu-i\Gamma_\nu/2$,
 where $\Gamma_\nu$ is the width of the resonance. 

  For the isotope $^{84}$Ni we have chosen the following Wood-Saxon 
 parameters: $V_0$=42.0 MeV, $V_{so}$=16.0 MeV, $r_0$=1.27 fm and
  $a$ = 0.67 fm. As in the previous case, the values of $r_0$ and $a$ 
 are the same for the mean field and for the spin-orbit interaction.
 The single-particle states corresponding to the major shell N=50-82
 are given in Table \ref{sps}. It can be seen that at variance to the
 situation in stable nuclei ( e.g., tin isotopes) the states
 $1d_{3/2}$, $1g_{7/2}$ and $1h_{11/2}$ are not bound states
 but single-particle resonances. In the presence of pairing correlations
 these states will generate  the quasiparticle resonances discussed in
 the subsection C.

\subsection{The BCS solution: contribution of resonant continuum} 

 With the single-particle states described in the previous section we have
 solved the BCS equations for the pairing force given by Eq.(5). For the
 parameters of the pairing force we have used the following values:
 a) oxygen isotopes: $V_0$=-456 MeV fm$^{-3}$, $\eta$=1, $\alpha$=1; b) nickel
isotope:
 $V_0$=-1130 MeV fm$^{-3}$, $\eta$=1, $\alpha$=1.In both cases we have introduced in
the
 calculations all the bound states. In order to analyse various approximations
 of continuum treatment, we have performed two types of BCS calculations.

 First, we have considered in the BCS equations the whole contribution of the
 continuum states up to a cut-off energy equal to 10 MeV. The continuum contribution
 is introduced through the scattering states of real energy. The number of scattering
 states per energy interval was increased, especially in the region of single-particle
 resonances, up to the convergence. In the
 calculations we have considered for the continuum states the same (lj) values
as for the
 bound states. These calculations are called below continuum-BCS (cBCS).

 In the second BCS calculation we have kept from the continuum only the contribution
 corresponding to the resonant states. More precisely, we have included in the calculations
 only the scattering states with the energies in the interval 
$\varepsilon_\nu \pm 2\Gamma_\nu$,  where $\varepsilon_\nu$ and $\Gamma_\nu$ are the
energies and the widths of the resonances 
 given in Table \ref{sps}. These calculations are refered to as resonant-BCS (rBCS)
 \cite{ni2}.

 \begin{figure}
      \includegraphics[width=0.5\textwidth]{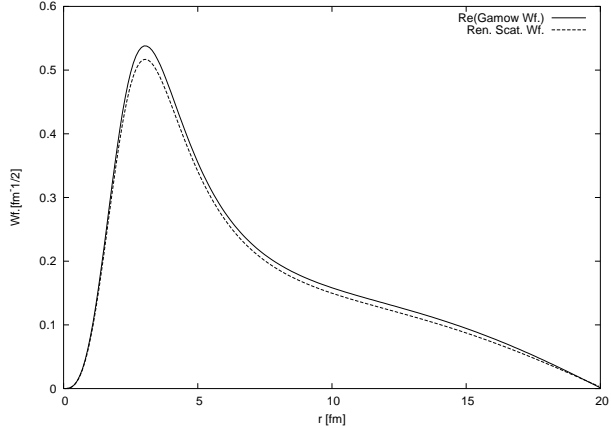}
 \caption{The scattering state $d_{3/2}$  evaluated at the position of the 
  $1d_{3/2}$ resonance, i. e. 0.910 MeV (full line). The dashed line shows
  the real part of the corresponding Gamow function, with the energy
  equal to (0.910,-0.050) MeV.
 \label{fig:rsf}}
 \end{figure}
   
 The results of these two calculations are shown in Table \ref{avg1}. In order to see
what
 is the total contribution of the continuum, in Table \ref{avg1} we give also the results
 obtained when in the calculations we include only the bound states (bBCS).

 \begin{table*}
 \caption{\label{avg1}
          Average gap, pairing energies (in MeV)
          and root-mean-square radii (in fm) calculated by using
          the representations discussed in the tex.}
 \begin{ruledtabular}
 \begin{tabular}{cccccccccccc}
          &          & cBCS  &       &$~~~$&          & rBCS  &      &$~~~$&
                     bBCS  &       \\
          &$<\Delta>$& $E_p$ & $<r>$ &$~~~$&$<\Delta>$& $E_p$ & $<r>$ &$~~~$&
           $<\Delta>$& $E_p$ & $<r>$ \\
 \hline
  $^{20}$O & 1.831 & -1.044 & 2.916 &$~~~$& 1.800 & -1.013 & 2.913 &$~~~$&
             1.784 &  -1.002 & 2.912   \\
  $^{22}$O & 1.412 & -0.876 & 3.040 &$~~~$& 1.384 & -0.856 & 3.038&$~~~$&
             1.370 & -0.849 & 3.038  \\
  $^{84}$Ni & 1.493 & -0.644 & 4.650 &$~~~$& 1.390 & -0.610 & 4.647 &$~~~$&
             1.386 & -0.606 & 4.646     
 \end{tabular}
 \end{ruledtabular}
 \end{table*}

 From Table \ref{avg1} one can first notice that for all the isotopes analysed here
 the contribution of the continuum to the correlation energies is rather small.
 Secondly, one can see that the contribution of the continuum is essentially
 given by a few low-lying single-particle resonances. In the presence of
 pairing correlations they generate the resonant quasiparticles states, 
 analysed in the next subsection.

\subsection{Quasiparticle resonances}

 The quasiparticle resonances, which describe unbound excited
 states in odd-even nuclei, are obtained  from the BCS equations
 written in Berggren representation. In this representation each
 quasiparticle resonance is described by a unique state, corresponding
 to a single-particle Gamow state. This is different from the case
 of real energy representations, where a quasiparticle resonance
 is charaterized by a continuous set of scattering states with energies
 in the vicinity of the resonance energy.

 As it was already discussed in the previous section, in order to get the
 energies and the widths of the quasiparticle resonances is necessary to
 calculate the pairing field and the chemical potential. Since these
 two quantities do not depends on the representation, we can calculate
 them by solving the BCS equations by using a real energy representation.
 Thus, we have solved the BCS equations by using all the bound states and
 the whole contribution of the continuum states up to a cut-off energy equal
 to 10 MeV. The continuum contribution was introduced through the scattering
 states of real energy. This calculation corresponds to the case cBCS 
 described in the previous subsection.

 With the mean field and the chemical potentials provided by the cBCS
 equations we then calculated, by using Eqs.(11c), the pairing gaps
associated
 to a Gamow state and the corresponding quasiparticle energies and widths.
 The results are shown in Table \ref{qpr}. From these results one can see
 clearly how the widths of single- particle resonances are changing by the
 pairing correlations. This direct accesss to the widths of quasiparticle
 resonances is the main advantage of expressing the BCS solution in
 terms of single-particle Gamow functions. 

\begin{turnpage}
 \begingroup
 \squeezetable
 \begin{table}
 \caption{\label{qpr}
          The quasiparticle energies, $E_i$,  the
          corresponding single-particle energies measured
          from the chemical potential,i.e., $\epsilon_i=|\varepsilon_i -\lambda|$
          and the occupation probabilities. The energies are in MeV.}
 \begin{ruledtabular}
 \begin{tabular}{ccccccccccc}
  State $i$ &              & $^{20}$O  &
            &              & $^{22}$O  &         &
  State $i$ &              & $^{84}$Ni &         \\
            & $\epsilon_i$ & $E_i$     & $v^2_i$
            & $\epsilon_i$ & $E_i$     & $v^2_i$ &
            & $\epsilon_i$ & $E_i$     & $v^2_i$ \\
 \hline
 $1d_{5/2}$ & (0.301,0)    & (1.656,0) & (0.591,0)
           & (1.060,0)    & (1.657,0) & (0.820,0) &
 $2d_{5/2}$ & (1.255,0)    & (1.720,0) & (0.864,0) \\
 $2s_{1/2}$ & (0.564,0)    & (1.301,0) & (0.283,0)
           & (0.195,0)    & (1.020,0) & (0.596,0) &
 $3s_{1/2}$ & (0.631,0)    & (1.044,0) & (0.801,0) \\
 $1d_{3/2}$ &(4.736,-0.141)&(4.878,-0.084)&(0.015,-0.003)
           &(3.977,-0.148)&(4.082,-0.075)&(0.013,-0.003) &
 $2d_{3/2}$ &(0.701,-0.050)&(1.050,-0.183)&(0.171,-0.030) \\
 $1f_{7/2}$ &(11.273,-1.528)&(11.310,-1.574)&(0.002,-0.001)
           &(10.514,-1.528)&(10.541,-1.559)&(0.001,-0.001) &
 $1f_{7/2}$ &(2.121,-0.010)&(2.486,-0.021)&(0.073,-0.002) \\
           &              &              &
           &              &              &                &
 $1h_{11/2}$ &(3.745,-0.018)&(3.937,-0.025)&(0.024,-0.001)
 \end{tabular}
 \end{ruledtabular}
 \end{table}
 \endgroup
\end{turnpage}

\section{Summary and conclusions} \label{sec:sum}
 In this paper we have discussed how the energies and the widths of
 quasiparticle resonances can be calculated by using the BCS
 equations written in Berggren representation. In this representation
 the pairing gaps associated to the  quasiparticle resonances, $\Delta_\nu$,
 are obtained by integrating the pairing field with the square of Gamow states.
 On the other hand, the pairing field, which does not depend on representation,
 is calculated by solving the BCS equations in a single-particle basis formed
 from bound states and scattering states of real energies. Then, the energies
 associated to the quasiparticle resonances $E_\nu$ are calculated by using the
 standard BCS  formula,i.e., $E_\nu=\sqrt{(\epsilon_\nu-\lambda)^2 +
 \Delta_\nu^2}$, where $\epsilon_\nu$ are the energies of the Gamow states.
 Since both $\epsilon_\nu$ and $\Delta_\nu$ are complex quantities,
 the quasiparticle energies are also complex. As in the case of single-particle
 resonances, the imaginary parts of $E_\nu$ are associated to the widths of
 the quasiparticle resonances.

 The calculation of quasiparticle resonances was illustrated for the case of
 the isotopes $^{20-22}$O and $^{84}$Ni. It was shown that in these isotopes
 the contribution of the single-particle resonant states upon pairing
 correlations is dominant compared to the contribution of background
 continuum. The most relevant aspect of the BCS calculations presented here
 is that they describe the excitations in resonant quasiparticle states
 in term of individual states, as in the case of bound states excitations.

\acknowledgments
 This work has been supported by FOMEC and Fundaci\'on Antorchas (Argentina), 
 by the Hungarian OTKA fund Nos. T37991 and T46791 and by the 
 Swedish Foundation for International Cooperation in Research and
 Higher Education (STINT).

\end{document}